\documentclass[a4paper]{article}

\usepackage{INTERSPEECH_v2}
\usepackage{dsfont}
\usepackage{graphicx}
\usepackage{xcolor}
\usepackage{caption}
\usepackage{subcaption}
\usepackage{multirow}
\usepackage{color,soul}

\title{Capturing Long-term Temporal Dependencies with Convolutional 
	   Networks for Continuous Emotion Recognition}
  
\name{Soheil Khorram$^{*1}$, Zakaria Aldeneh$^{*1}$, Dimitrios Dimitriadis$^2$, 
      \\Melvin McInnis$^{1}$, Emily Mower Provost$^{1}$}
\address{$^1$University of Michigan at Ann Arbor, $^2$IBM T. J. Watson Research Center}
\email{\texttt{\{khorrams, aldeneh, mmcinnis, emilykmp\}@umich.edu,
        dbdimitr@us.ibm.com}}
\begin{document}

\maketitle

%%%%%%%%%%%%%%%%%%%%%%%%%%%%%%%%%%%%%%%%%%%%%%%%%%%%%%%%%%%%%%%%%%%%%%%%%%%%%%%
%%%%%%%%%%%%%%%%%%%%%%%%%%%%%%%%%%%%%%%%%%%%%%%%%%%%%%%%%%%%%%%%%%%%%%%%%%%%%%%

\begin{abstract}
  The goal of continuous emotion recognition is to assign an emotion value to 
  every frame in a sequence of acoustic features.
  We show that incorporating long-term temporal dependencies is critical
  for continuous emotion recognition tasks. To this end, we first investigate 
  architectures that use dilated convolutions. We show that even though such
  architectures outperform previously reported systems, the output signals 
  produced from such architectures undergo erratic changes between
  consecutive time steps. This is inconsistent with the slow moving 
  ground-truth emotion labels that are obtained from human annotators. 
  To deal with this problem, we model a downsampled version of the input signal
  and then generate the output signal through upsampling. Not only
  does the resulting downsampling/upsampling network achieve good performance,
  it also generates smooth output trajectories.
  Our method yields the best known audio-only performance on the RECOLA
  dataset.
  
\end{abstract}
  
\makeatletter{\renewcommand*{\@makefnmark}{}
	\footnotetext{$^*$These authors contributed equally to this work}}

%%%%%%%%%%%%%%%%%%%%%%%%%%%%%%%%%%%%%%%%%%%%%%%%%%%%%%%%%%%%%%%%%%%%%%%%%%%%%%%
%%%%%%%%%%%%%%%%%%%%%%%%%%%%%%%%%%%%%%%%%%%%%%%%%%%%%%%%%%%%%%%%%%%%%%%%%%%%%%%

\section{Introduction}
  Emotion recognition has many potential applications including
  building more natural human-computer interfaces.
  Emotion can be quantified using categorical classes (e.g., \textit{neutral,
  happy, sad, etc.}) or using dimensional values (e.g., \textit{valence-arousal}).
  In addition, emotional labels can be quantified statically, over units of 
  speech (e.g., utterances), or continuously in time.

  In this work, we focus on problems where the goal is to recognize emotions
  in the valence-arousal space, continuously in time.
  The valence-arousal space is a psychologically grounded method for describing 
  emotions~\cite{lewis2010handbook}. Valence ranges from negative to positive, 
  while activation ranges from calm to excited.  
  Research has demonstrated that it is critical to incorporate 
  long-term temporal information for making accurate emotion predictions.
  For instance,
  Valstar et al.~\cite{valstar2016avec} showed that it was necessary to
  consider larger windows when making frame-level emotion predictions 
  (four seconds for arousal and six seconds for valence).
  Le et al.~\cite{le2013emotion} and Cardinal et al.~\cite{cardinal2015ets} 
  found that increasing the number of contextual frames when training a
  deep neural network (DNN) for making frame-level emotion predictions
  is helpful but only to a certain point.
  Bidirectional long short-term memory networks 
  (BLSTMs) can naturally
  incorporate long-term temporal dependencies between features;
  explaining their success in continuous emotion recognition tasks
  (e.g.,~\cite{he2015multimodal}).
  
  In this work, we investigate two convolutional network architectures,
  dilated convolutional networks and downsampling/upsampling networks, 
  that capture long-term temporal dependencies. 
  We interpret the two architectures in the context of continuous 
  emotion recognition and show that these architectures can be used to
  build accurate continuous emotion recognition systems.

%%%%%%%%%%%%%%%%%%%%%%%%%%%%%%%%%%%%%%%%%%%%%%%%%%%%%%%%%%%%%%%%%%%%%%%%%%%%%%%
%%%%%%%%%%%%%%%%%%%%%%%%%%%%%%%%%%%%%%%%%%%%%%%%%%%%%%%%%%%%%%%%%%%%%%%%%%%%%%%

\section{Related Work}
  Even though the problem of emotion recognition has been extensively studied 
  in the literature, we only focus on works that predicted 
  dimensional values, continuously in time.
  Successful attempts to solving the continuous emotion recognition problem 
  relied on DNNs~\cite{cardinal2015ets},
  BLSTMs~\cite{he2015multimodal}, and more commonly, support vector 
  regression (SVR) classifiers~\cite{brady2016multi}.
  With the exception of BLSTMs, such approaches do not incorporate
  long-term dependencies unless coupled with feature engineering. 
  In this work, we show that purely convolutional neural networks 
  can be used 
  to incorporate long-term dependencies and achieve good emotion recognition
  performance, and are more efficient to train than their recurrent
  counterparts.
  
  In their winning submission to the AVEC 2016 challenge, 
  Brady et al.~\cite{brady2016multi} extracted a set of audio features
  (Mel-frequency cepstral coefficients, shifted delta cepstral, 
  prosody) and then learned higher-level representations of the
  features using sparse coding. The higher-level audio features were
  used to train linear SVRs.
  Povolny et al.~\cite{povolny2016multimodal} used eGeMAPS
  \cite{eyben2016geneva} features along 
  with a set 
  of higher-level bottleneck features extracted from a DNN trained for 
  automatic speech recognition (ASR) to train linear regressors.
  The higher level features were produced from an initial set of $24$ 
  Mel filterbank (MFB) features
  and four different estimates of the fundamental frequency (F0). 
  Povolny et al. used all features to train linear regressors to
  predict a value for each frame, and considered two methods for 
  incorporating contextual information: simple frame stacking and
  temporal content summarization by applying statistics to local windows.
  In contrast, in this work we show that considering temporal dependencies 
  that are longer than those presented in 
  ~\cite{brady2016multi,povolny2016multimodal} is 
  critical to improve continuous emotion recognition performance.
  
  He et al.~\cite{he2015multimodal} extracted a comprehensive set of 
  $4,684$ features, which included energy, spectral, and voicing-related 
  features, and used them to train BLSTMs.
  The authors introduced delay to the input to compensate for human evaluation 
  lag and then applied feature
  selection. The authors ran the predicted time series through
  a Gaussian smoothing filter to produce the final output. 
  In this work, we show that it is sufficient to use $40$ MFBs to achieve
  state-of-the-art performance, without the need for
  special handling of human evaluation lag.
  
  Trigeorgis et al.~\cite{trigeorgis2016adieu} trained a convolutional 
  recurrent network for continuous emotion recognition using the time domain 
  signal directly. The authors split the utterances into five-second
  segments for batch training. 
  Given an output from a the 
  trained model, the authors applied a chain of post-processing steps 
  (median filtering, centering, scaling, time shifting) to get the 
  final output. 
  In contrast, we show that convolutional networks make it possible to
  efficiently process full utterances without the need for segmenting. 
  Further, since our models work on full-length
  utterances, we show that it is not necessary to apply any 
  post-processing steps as described in ~\cite{trigeorgis2016adieu}.
  
%   A number of problems can be viewed as dense prediction tasks,
%   where the goal is to assign a value to every frame or pixel in a
%   given input,
%   making it possible to utilize tools and techniques presented in other fields.
%   For instance, Sercu et al.~\cite{sercu2016dense} proposed viewing ASR
  On the ASR end, Sercu et al.~\cite{sercu2016dense} proposed viewing ASR 
  problems as dense prediction tasks,
  where the goal is to assign a label to every frame in a given sequence,
  and showed that this view provides a 
  set of tools (e.g., dilated convolutions, batch normalization, efficient 
  processing) that can improve ASR performance. The authors argued that 
  ASR approaches required practitioners to splice their
  input sequences into independent windows, making the training and evaluation
  procedures cumbersome and computationally inefficient. In contrast, the
  authors' proposed approach allows practitioners to efficiently process 
  full sequences without requiring splicing or processing frames independently.
  The authors showed that their approach obtained the best 
  published single model results on the switchboard-2000 benchmark dataset.
  
  In this work, we treat the problem of continuous emotion
  recognition as a dense prediction task and show that, given this
  view of the problem, we can utilize convolutional architectures
  that can efficiently incorporate long-term temporal dependencies
  and provide accurate emotion predictions. 

%%%%%%%%%%%%%%%%%%%%%%%%%%%%%%%%%%%%%%%%%%%%%%%%%%%%%%%%%%%%%%%%%%%%%%%%%%%%%%%
%%%%%%%%%%%%%%%%%%%%%%%%%%%%%%%%%%%%%%%%%%%%%%%%%%%%%%%%%%%%%%%%%%%%%%%%%%%%%%%

\section{Problem Setup}
  We focus on the RECOLA database ~\cite{ringeval2013introducing}
  following the AVEC 2016 guidelines~\cite{valstar2016avec}.
  The RECOLA database consists of spontaneous interactions in French and 
  provides continuous, dimensional (valence and arousal) ground-truth
  descriptions of emotions. Even though the AVEC 2016 challenge
  is multi-modal in nature, we only focus on the speech modality in this work. 
  The RECOLA database contains a total of $27$ five-minute utterances, each
  from a distinct speaker ($9$ train; $9$ validation; $9$ test).
  Ground-truth continuous annotations were computed,
  using audio-visual cues, on a temporal 
  granularity of $40$ms from six annotators (three females).

  \textbf{Features.} We use the Kaldi toolkit~\cite{povey2011kaldi} to extract
  40-dimensional log MFB features, using a window length of
  $25$ms with a hop size of $10$ms. 
  Previous work showed that MFB features are better than conventional MFCCs
  for predicting emotions~\cite{busso2007using}.
  We perform speaker-specific $z$-normalization on all extracted features. 
  RECOLA provides continuous labels at a granularity of $40$ms. Thus, we stack
  four subsequent MFB frames to ensure correspondence between hop sizes 
  in the input and output sequences.

  \textbf{Problem Setup.} Given a sequence of stacked acoustic features
  $ \bm{X} = [\bm{x}_1, \bm{x}_2, \dots, \bm{x}_T]$,
  where $\bm{x}_t \in \mathds{R}^{d}$,
  the goal is to produce a sequence of continuous emotion labels
  $\bm{y} = [y_1, y_2, \dots, y_T ]$, where $y_t \in \mathds{R}$.
  % In other words, the goal is to learn a mapping $f$ such that
  % $ f:\bm{X} \rightarrow \bm{y}$.
  
  \textbf{Evaluation Metrics.} Given a sequence of ground-truth labels 
  $\bm{y} = [y_1, y_2, \dots, y_T ]$ and a sequence of predicted labels
  $\bm{\hat{y}} = [\hat{y}_1, \hat{y}_2, \dots, \hat{y}_T]$, we evaluate
  the performance using the root mean squared error (RMSE) and the 
  Concordance Correlation Coefficient (CCC) to be consistent with previous work.
  The CCC is computed as follows:\\
  $\text{CCC} = 2\sigma^2_{y\hat{y}}/
  			   (\sigma^{2}_{y} + \sigma^{2}_{\hat{y}} + 
               (\mu_{y}-\mu_{\hat{y}})^2)$,
%   $$\text{CCC} = \frac{2\sigma^2_{y\hat{y}}}
%   			   {\sigma^{2}_{y} + \sigma^{2}_{\hat{y}} + 
%                (\mu_{y}-\mu_{\hat{y}})^2}$$
  where $\mu_{y} = \mathds{E}(\bm{y})$, 
  $\mu_{\hat{y}} = \mathds{E}(\bm{\hat{y}})$, 
  $\sigma^{2}_{y} = \text{var}(\bm{y})$,
  $\sigma^{2}_{\hat{y}} = \text{var}(\bm{\hat{y}})$, and
  $\sigma^2_{y\hat{y}} = \text{cov}(\bm{y},\bm{\hat{y}})$.

%%%%%%%%%%%%%%%%%%%%%%%%%%%%%%%%%%%%%%%%%%%%%%%%%%%%%%%%%%%%%%%%%%%%%%%%%%%%%%%
%%%%%%%%%%%%%%%%%%%%%%%%%%%%%%%%%%%%%%%%%%%%%%%%%%%%%%%%%%%%%%%%%%%%%%%%%%%%%%%
\section{Preliminary Experiment}
\begin{figure}[t]
      \centering
      \def\factor{0.6}
      \includegraphics[width=\factor\linewidth]{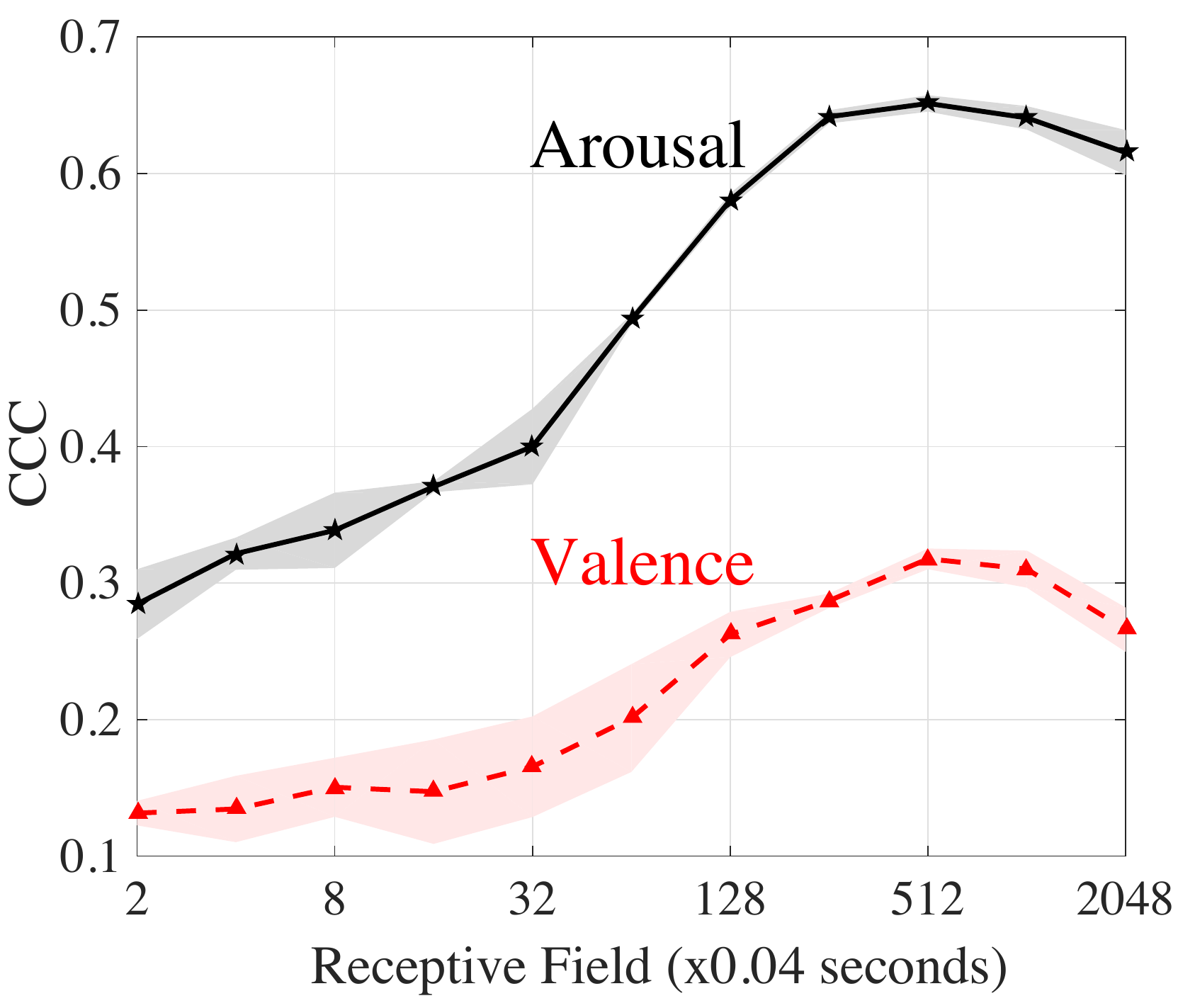}
      \caption{Increasing the size of the receptive field improves
      			performance for both arousal and valence. 
                Solid lines represent mean CCC from $10$ runs and shaded area
                represents standard deviation from the runs.}
      \label{fig:receptive_field_exp}
    \end{figure}

  We first study the effect of incorporating temporal dependencies of different
  lengths.
  The network that we use in the preliminary experiments 
  consists of a
  convolutional layer with one filter of variable length from $2$ to $2048$ 
  frames, followed by a $tanh$
  non-linearity, followed by a linear regression layer. We vary the length of 
  the filter and validate the performance using CCC.
  We train our model 
  on the training partition and evaluate on the development partition.
  We report the results of our preliminary experiment
  in Figure~\ref{fig:receptive_field_exp}.
  The results show that incorporating long-term 
  temporal dependencies improves the performance on the validation set up to 
  a point.
  
  The observed diminishing gains in 
  performance past $512$ ($20.48$ seconds) frames may occur either due to 
  the increased number of parameters or because contextual information 
  becomes irrelevant after $512$ frames.
  Covering contexts as large as $512$ frames still provided 
  improvements in performance compared to results obtained from covering 
  smaller contexts.	
  The utility of contexts spanning $512$ frames ($20.48$ seconds)
  is contrary to previous work that considered much smaller time
  scales.
  For instance, Valstar et al.~\cite{valstar2016avec} only covered six 
  seconds worth of features and Povolny et al.~\cite{povolny2016multimodal}
  considered a maximum of eight seconds worth of features. 
  Results from the preliminary experiment suggest that continuous emotion
  prediction systems could
  benefit from incorporating long-term temporal dependencies. This acts as a
  motivation for using architectures that are specifically designed for
  considering long-term dependencies.

%%%%%%%%%%%%%%%%%%%%%%%%%%%%%%%%%%%%%%%%%%%%%%%%%%%%%%%%%%%%%%%%%%%%%%%%%%%%%%%
%%%%%%%%%%%%%%%%%%%%%%%%%%%%%%%%%%%%%%%%%%%%%%%%%%%%%%%%%%%%%%%%%%%%%%%%%%%%%%%

\section{Methods}
  In this section, we describe the two architectures that we propose to use
  to capture long-term temporal dependencies in continuous emotion prediction
  tasks.

  \subsection{Dilated Convolutions}
  	Dilated convolutions provide an efficient way
    to increase the receptive field without causing the number of learnable
    parameters to vastly increase. Networks that use dilated convolutions
    have shown success in a number of tasks, including
    image segmentation ~\cite{yu2015multi}, speech synthesis
    \cite{van2016wavenet} and ASR~\cite{sercu2016dense}.

    van den Oord et al.~\cite{van2016wavenet} recently showed that it is 
    possible to use convolutions with various dilation factors to allow the
    receptive field of a generative model to grow exponentially in order 
    to cover thousands of time steps and synthesize high-quality speech.
    Sercu et al.~\cite{sercu2016dense} showed that ASR could 
    benefit from dilated convolutions since they allow larger
    regions to be covered without disrupting the length of the input signals.
    Continuous emotion recognition could benefit from such properties.
    
\begin{figure}[t]
	  \centering
	  \def\factor{0.7}
      \includegraphics[width=\factor\linewidth]{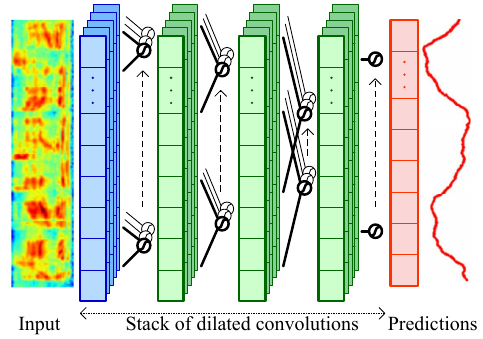}
      \caption{A visualization of our dilated convolution network. We use 
      		   convolutions with a different dilation factor for different layers.
               We use a $1\times1$ convolution for the last layer to produce the 
               final output.}
      \label{fig:dilated}
    \end{figure}

    When compared to filters of regular convolutions, those of dilated 
    convolutions touch
    the input signal every $k$ time steps, where $k$ is the 
    \textit{dilation factor}. If $[w_1,w_2,w_3]$ is a filter with a dilation
    factor of zero, then $[w_1, 0, w_2, 0, w_3]$ is the filter with a dilation
    factor of one and $[w_1, 0, 0, w_2, 0, 0, w_3]$ is the filter with a 
    dilation factor of two, and so on.
    We build a network that consists of stacked convolution layers,
    where the convolution functions in each layer use a dilation factor
    of $2^n$, where $n$ is the layer number. This causes the dilation factors
    to grow exponentially with depth while the number of parameters grows
    linearly with depth. Figure~\ref{fig:dilated} shows a diagram of our
    dilated convolution network.
    
    \begin{figure}[t]
	  \centering
	  \def\factor{0.8}
      \includegraphics[width=\factor\linewidth]{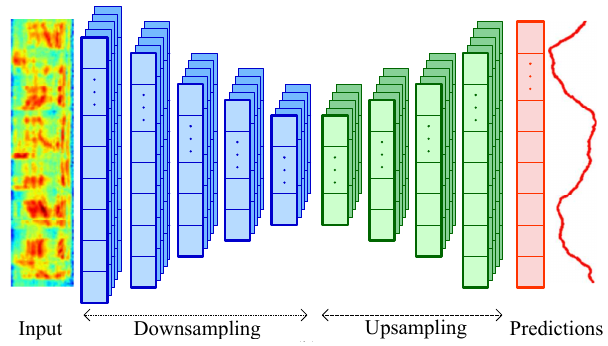}
      \caption{A visualization of our downsampling/upsampling network.
      		   Downsampling compresses the input signal into shorter signal which
               is then used to reconstruct a signal of the same length by the
               upsampling sub-network. We use the transpose convolution operation
               to perform upsampling.}
      \label{fig:deconv}
    \end{figure}

  \subsection{Downsampling/Upsampling}
  	The emotion targets in the RECOLA database are sampled at a frequency of 
    $25$ Hz. 
    Using Fourier analysis, we find that more than $95$ percent of the power 
    of these trajectories lies in frequency bands that are lower than $1$ Hz.
    In other words, the output signals are smooth and they have considerable
    time dependencies. This finding is not surprising because we do not expect
    rapid reactions from human annotators.
    Networks that use dilated convolutions do not take
    this fact into account while making predictions, causing them to generate
    output signals whose variance is not consistent with the continuous ground
    truth contours (Section~\ref{Results}). To deal with this problem, 
    we propose the use of
    a network architecture that compresses the input signal into a 
    low-resolution signal through downsampling
    and then reconstructs the output signal through upsampling. Not only
    does the downsampling/upsampling architecture capture long-term temporal
    dependencies, it also generates a smooth output trajectory.
    
\begin{figure}[t]
      \centering
      \def\factor{0.6}
      \includegraphics[width=\factor\linewidth]{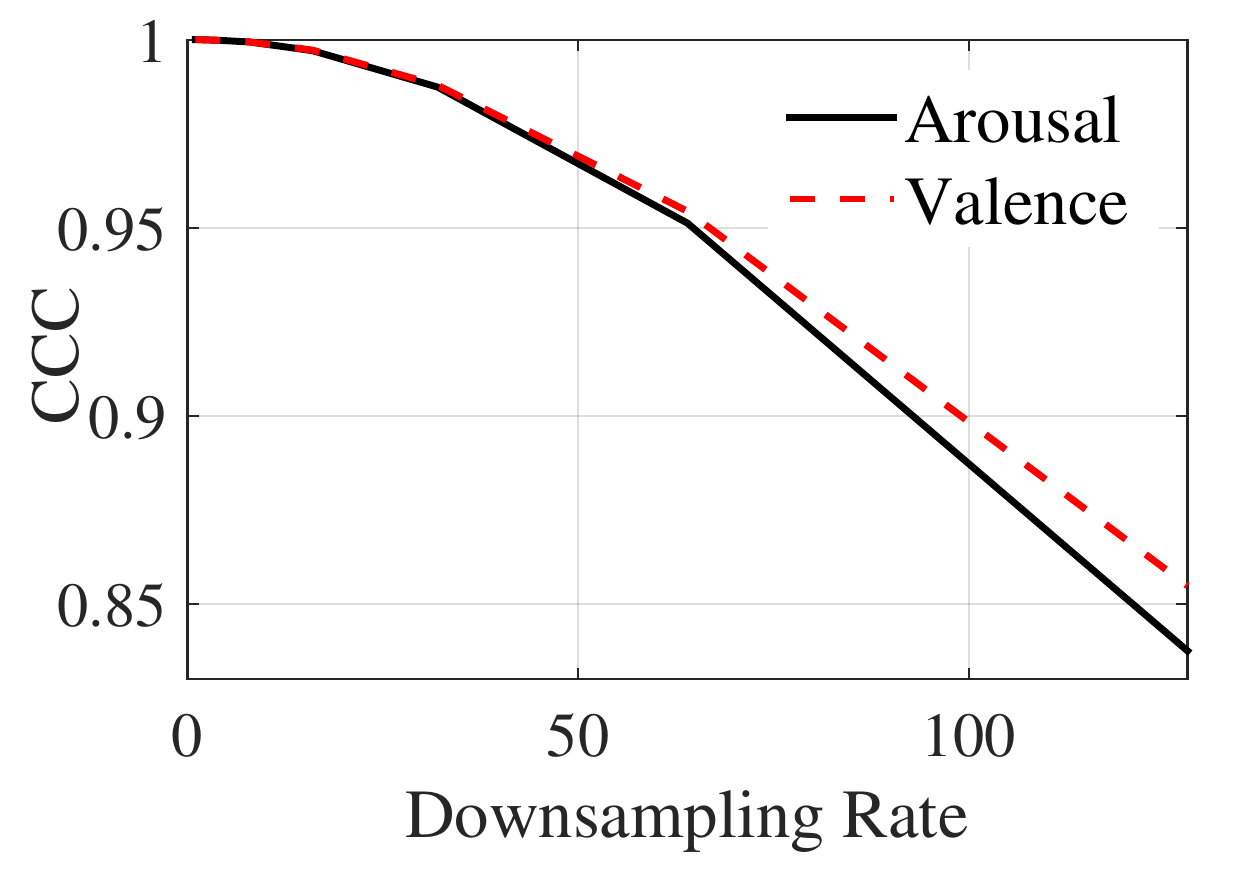}
      \caption{Effect of downsampling/upsampling on CCC.}
      \label{fig:sampling_exp}
    \end{figure}

    We conduct an experiment to investigate the effect of 
    downsampling/upsampling 
    on continuous emotion labels. First, we convert the ground truth 
    signals to low-resolution signals using standard uniform downsampling. 
    Given the downsampled signals, we then generate the original signals 
    using spline interpolation. We vary the downsampling factor exponentially
    from $2$ to $128$ and compute the CCC between the original signals and the
    reconstructed ones. The results that we show in Figure~\ref{fig:sampling_exp}
    demonstrate that distortions caused by downsampling with factors up to $64$ 
    are minor ($<5\%$ loss in CCC relative to original). 
        
    The network that we use contains two subnetworks: (1) a downsampling network;
    (2) an upsampling network. The downsampling network consists of a series of
    convolutions and max-pooling operations. The max-pooling layers reduce the 
    resolution of the signal and increase the effective receptive field of
    the convolution layers. Initial experiments showed that max-pooling was
    more effective than other pooling techniques.
    
    The upsampling function can be implemented in a number of ways 
    \cite{noh2015learning}. 
    In this work we use the transposed convolution\footnote{Other names in 
    literature include deconvolution, upconvolution, backward strided 
    convolution and fractionally strided convolution.} 
    \cite{zeiler2010deconvolutional, dumoulin2016guide} operation to perform upsampling.
    Transposed convolutions provide a learnable map that can upsample a 
    low-resolution signal to a high-resolution one. In contrast to standard
    convolution filters that connect multiple input samples to a single output 
    sample, transposed convolution filters generate multiple outputs samples 
    from just one input sample. Since it generates multiple outputs 
    simultaneously, the transposed convolution can be thought of as
    a learnable interpolation function.
    
    Downsampling/upsampling architectures have been used in many computer 
    vision tasks (e.g.,~\cite{noh2015learning, badrinarayanan2015segnet, 
    chen2016single}). 
    For instance, Noh et al.~\cite{noh2015learning} showed that transposed 
    convolution operations can be effectively applied to image segmentation 
    tasks. 
    In addition to vision applications, downsampling/upsampling architectures
    have been successfully applied to speech enhancement 
    problems~\cite{park2016fully}, where the goal is to learn a mapping between
    noisy speech spectra and their clean counterparts. 
    Park et al.~\cite{park2016fully} demonstrated that downsampling/upsampling
    convolutional networks can be $12\times$ smaller (in terms of the number
    of learnable parameters) than their recurrent counterparts and yet yield 
    better performance on speech enhancement tasks.
   
    The main goal of a transposed convolution is to take an $n_x$-dimensional
    low-resolution vector $\bm{x}$ and generate an $n_y$-dimensional 
    high-resolution vector $\bm{y}$ using an $n_w$-dimensional filter $\bm{w}$ (where $n_y > n_x$).
    Similar to other linear transforms, $\bm{y}$ can be expressed as: 
    $\bm{y} = \bm{Tx}$, where $\bm{T}$ is the linear $n_y$-by-$n_x$ transform
    matrix that is given by
    $\bm{T} = [\bm{T}_1, \bm{T}_2, ..., \bm{T}_{n_x}]$. $\bm{T}_i$ is the $i$-th column of $\bm{T}$ and can be written as:
    \[\bm{T}_i = [\underbrace{0,...,0}_{s(i-1)}, \underbrace{\bm{w}^T}_{n_w},
       \underbrace{0,...,0}_{s(n_x-i)}]^T \]
    where $s$ is the upsampling factor. This linear interpolator is able to 
    expand the input vector $\bm{x}$ to the output vector $\bm{y}$ with the length
    of $n_y=s(n_x-1)+n_w$. Note that the matrix $\bm{T}$ is nothing but the transposed
    version of the standard strided convolution transform matrix.
    Our experiments confirm that the proposed 
    downsampling/upsampling network generates smooth trajectories. 

%%%%%%%%%%%%%%%%%%%%%%%%%%%%%%%%%%%%%%%%%%%%%%%%%%%%%%%%%%%%%%%%%%%%%%%%%%%%%%%
%%%%%%%%%%%%%%%%%%%%%%%%%%%%%%%%%%%%%%%%%%%%%%%%%%%%%%%%%%%%%%%%%%%%%%%%%%%%%%%

	\begin{figure}[t]
      \centering
      \def\factor{0.7}
      \includegraphics[width=\factor\linewidth]{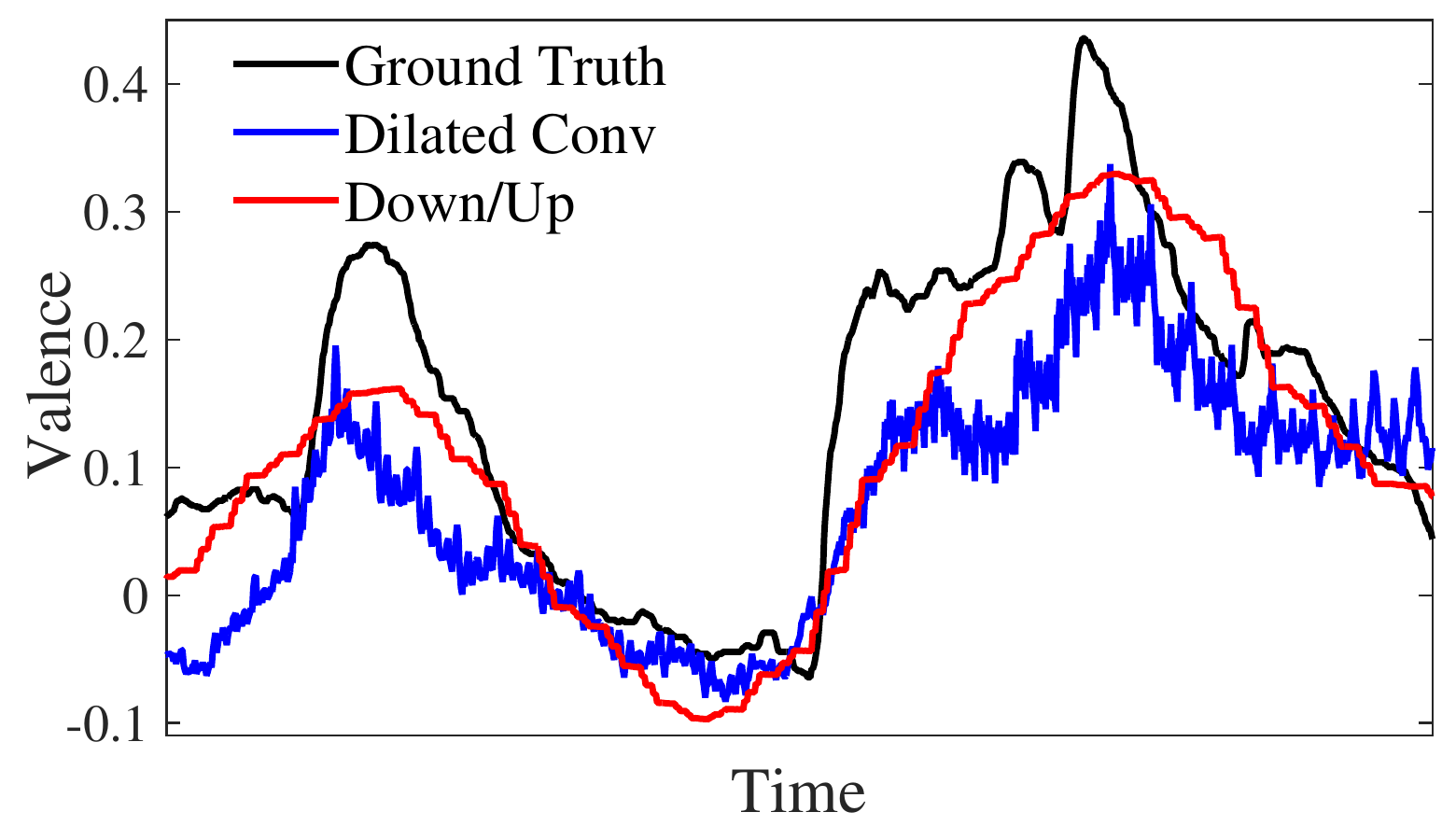}
      \caption{A visualization of the predictions produced by the two models
      			plotted against ground-truth for a $40$-second segment.}
      \label{fig:smoothness}
    \end{figure}

\section{Results and Discussion}
  \subsection{Experimental Setup}
  	We build our models using Keras~\cite{chollet2015keras} with a 
    Theano backend~\cite{bergstra2010theano}. We train our models on the 
    training partition of the dataset and use the development partition
    for early stopping and hyper-parameter selection (e.g., learning rate, 
    number of layers layer size, filter width, $l2$ regularization, 
    dilation factors, downsampling factors). We optimize CCC
    directly in all setups.
    We repeat each experiment five times to account for the effect of 
    initialization.
    The final test evaluation is done by the
    AVEC 2016 organizers (i.e., we do not have access to test labels). Our test
    submissions were created by averaging the predictions produced from the
    five runs.
    
    We report published results from the literature as baselines. 
    Almost all previous works only 
    report their final test results based on multi-modal features.
    We only show results that are reported on the audio modality in the results
    tables. We also compare our performance to that of an optimized BLSTM 
    regression model, described in~\cite{le_discrete}.
    Our final dilated convolution structure has a depth of $10$ layers,
    each having a width of $32$. Our final downsampling/upsampling
    network contains four downsampling layers, one intermediate layer,
    and four transposed convolution layers, each having width of $32$ for
    arousal and $128$ for valence. We use a downsampling factor of three.
    We do not splice the input utterances into segments. Instead, we
    train on full length utterances and use a batch size of one.
    
  \subsection{Results}\label{Results}
  	Tables~\ref{tab:arousal_results} and~\ref{tab:valence_results} show the
    development and test results for arousal and valence, respectively.
    Each row shows the results for one setup.
    We only include results from the literature that are based on the speech
    modality and use ``--'' to show unreported results.
    
    Both proposed systems show improvements over baseline results by 
    Valstar et al.~\cite{valstar2016avec}. Our dilated convolution based 
    system provides improvements of $5.6\%$ and $19.5\%$ over baseline systems
    for arousal and valence, respectively. Our downsampling/upsampling
    system provides improvements of $5.1\%$ and $33.9\%$ over baseline systems
    for arousal and valence, respectively.
    We report the results we obtain from our BLSTM system to provide a
    reference point. Our BLSTM system performs well when
    compared to the baseline results. 
   
    The proposed methods outperform BLSTMs and are more efficient
    to train on long utterances. For instance, given a convolutional network 
    and a BLSTM network with approximately equal number of learnable parameters, 
    one epoch of training on the AVEC dataset takes about $13$ seconds
    on the convolutional network while one epoch of training takes about $10$
    minutes on the BLSTM network. This suggests that convolutional
    architectures can act as replacement for recurrent ones for continuous
    emotion recognition problems.
 
    \begin{table}[t]
      \caption{Arousal results.}
      \label{tab:arousal_results}
      \centering
      \begin{tabular}{ccccc}
      \toprule
      \multicolumn{1}{c}{\multirow{2}{*}{\textbf{Method}}} & 
      \multicolumn{2}{c}{\textbf{Dev.}} & 
      \multicolumn{2}{c}{\textbf{Test}} \\
      \multicolumn{1}{c}{} & RMSE & CCC & RMSE & CCC \\
      \midrule
          Valstar et al.~\cite{valstar2016avec} &  -- & $.796$ & -- & $.648$ \\
          Brady et al.~\cite{brady2016multi} & $.107$ & $.846$ & -- & -- \\
          Povolny et al.~\cite{povolny2016multimodal}$^{*}$ & $.114$ & $.832$ & $.141$ & $.682$ \\
          BLSTM~\cite{le_discrete}& $.103$ & $.853$ & $.143$ & $.664$\\
          Dilated& $.102$ & $.857$ & $\textbf{.137}$ & $\textbf{.684}$\\
          Down/Up& $\textbf{.100}$ & $\textbf{.867}$ & $\textbf{.137}$ & $.681$\\
          \bottomrule
      \end{tabular}
	\end{table}
	\begin{table}[t]
      \caption{Valence results.}
      \label{tab:valence_results}
      \centering
      \begin{tabular}{ccccc}
      \toprule
      \multicolumn{1}{c}{\multirow{2}{*}{\textbf{Method}}} & 
      \multicolumn{2}{c}{\textbf{Dev.}} & 
      \multicolumn{2}{c}{\textbf{Test}} \\
      \multicolumn{1}{c}{} & RMSE & CCC & RMSE & CCC \\
      \midrule
          Valstar et al.~\cite{valstar2016avec} &  -- & $.455$ & -- & $.375$ \\
          Brady et al.~\cite{brady2016multi} & $.132$ & $.450$ & -- & -- \\
          Povolny et al.~\cite{povolny2016multimodal}$^{*}$ & $.142$ & $.489$ & $.355$ & $.349$ \\
          BLSTM~\cite{le_discrete}& $.113$ & $.518$ & $\textbf{.116}$ & $.499$\\
          Dilated& $.117$ & $.538$ & $.121$ & $.486$\\
          Down/Up& $\textbf{.107}$ & $\textbf{.592}$ & $.117$ & $\textbf{.502}$\\
          \bottomrule
      \end{tabular}
	\end{table}
 
    We show an example $40$-second segment of the predictions made by our two
    networks along with the ground-truth predictions in Figure~\ref{fig:smoothness}.
    The figure shows that the predictions produced by the downsampling/upsampling
    network are much smoother than those produced by the dilated
    convolution networks. We believe that the structure of the downsampling/upsampling
    network forces the output to be smooth by generating the output from a compressed
    signal. The compressed signal only stores essential information that is necessary
    for generating trajectories, removing any noise components.
    
    \makeatletter{\renewcommand*{\@makefnmark}{}
	\footnotetext{$^*$Unpublished test results, courtesy of the authors.}}

%%%%%%%%%%%%%%%%%%%%%%%%%%%%%%%%%%%%%%%%%%%%%%%%%%%%%%%%%%%%%%%%%%%%%%%%%%%%%%%
%%%%%%%%%%%%%%%%%%%%%%%%%%%%%%%%%%%%%%%%%%%%%%%%%%%%%%%%%%%%%%%%%%%%%%%%%%%%%%%

\section{Conclusion}
  We investigated two architectures that provide different means for capturing 
  long-term temporal dependencies in a given sequence of acoustic features.
  Dilated convolutions provides a method for incorporating long-term 
  temporal information without disrupting the length of the input signal by
  using filters with varying dilation factors.
  Downsampling/upsampling networks incorporate long-term dependencies by 
  applying a series of convolutions and max-poolings to downsample the signal
  and get a global view of the features. The downsampled signal is then used
  to reconstruct an output with a length that is equal to the uncompressed
  input. Our methods achieve the best known audio-only performance on the 
  AVEC 2016 challenge.
  
\section{Acknowledgement}
This work was partially supported by IBM under the Sapphire project. We would like to thank Dr. David Nahamoo and Dr. Lazaros Polymenakos, IBM Research, Yorktown Heights, for their support.
  
%%%%%%%%%%%%%%%%%%%%%%%%%%%%%%%%%%%%%%%%%%%%%%%%%%%%%%%%%%%%%%%%%%%%%%%%%%%%%%%
%%%%%%%%%%%%%%%%%%%%%%%%%%%%%%%%%%%%%%%%%%%%%%%%%%%%%%%%%%%%%%%%%%%%%%%%%%%%%%%

% \section{Acknowledgements}

% The ISCA Board would like to thank the organizing committees of the past
% INTERSPEECH conferences for their help and for kindly providing the template
% files.

%%%%%%%%%%%%%%%%%%%%%%%%%%%%%%%%%%%%%%%%%%%%%%%%%%%%%%%%%%%%%%%%%%%%%%%%%%%%%%%
%%%%%%%%%%%%%%%%%%%%%%%%%%%%%%%%%%%%%%%%%%%%%%%%%%%%%%%%%%%%%%%%%%%%%%%%%%%%%%%

\bibliographystyle{IEEEtran}

\bibliography{mybib}

\end{document}